\documentclass[twocolumn,showpacs,aps,prl,superscriptaddress,floatfix]{revtex4}
\usepackage{graphicx}
\usepackage{citesort}
\usepackage{epic,eepic}
\usepackage{graphics}
\usepackage{epsfig}
\usepackage{subfigure}

\newcommand{\be}{\begin{equation}}
\newcommand{\ee}{\end{equation}}
\usepackage{amsmath}
\usepackage{amsfonts}
\usepackage{amssymb}
\usepackage{bm}
\usepackage{color}
\usepackage{graphicx}
\begin{document}
\title{Spontaneous Breakdown of Superhydrophobicity}
\author{Mauro Sbragaglia}\affiliation{Physics of Fluids, Department of
Applied Physics, University of Twente, Faculty of Science and Technology, Impact and Mesa$^{+}$ Institutes, P.O. Box 217, 7500 AE Enschede, The Netherlands.}
\author{Alisia M. Peters}\affiliation{Membrane Technology Group, Department 
of Chemical Engineering,  University of Twente, Faculty of Science and Technology, Impact and Mesa$^{+}$ Institutes, P.O. Box 217, 7500 AE Enschede, The Netherlands.}  
\author{Christophe Pirat} \affiliation{Physics of Fluids, Department of Applied Physics, University of Twente, Faculty of Science and Technology, Impact and Mesa$^{+}$ Institutes, P.O. Box 217, 7500 AE Enschede, The Netherlands.}
\author{Bram M. Borkent}\affiliation{Physics of Fluids, Department of Applied Physics, University of Twente, Faculty of Science and Technology, Impact and Mesa$^{+}$ Institutes, P.O. Box 217, 7500 AE Enschede, The Netherlands.}
\author{Rob G. H. Lammertink}\affiliation{Membrane Technology Group, Department of Chemical Engineering,  University of Twente, Faculty of Science and Technology, Impact and Mesa$^{+}$ Institutes, P.O. Box 217, 7500 AE Enschede, The Netherlands.}   
\author{Matthias Wessling}\affiliation{Membrane Technology Group, Department of Chemical Engineering,  University of Twente, Faculty of Science and Technology, Impact and Mesa$^{+}$ Institutes, P.O. Box 217, 7500 AE Enschede, The Netherlands.}  
\author{Detlef Lohse}\affiliation{Physics of Fluids, Department of Applied Physics, University of Twente, Faculty of Science and Technology, Impact and Mesa$^{+}$ Institutes, P.O. Box 217, 7500 AE Enschede, The Netherlands.}
\date{\today}
\begin{abstract}
In some cases water droplets can completely wet micro-structured superhydrophobic surfaces. The {\it dynamics} of this rapid process is analyzed by ultra-high-speed imaging. Depending on the scales of the micro-structure, the wetting fronts propagate smoothly and circularly or -- more interestingly -- in a {\it stepwise} manner, leading to a growing {\it square-shaped} wetted area: entering a new row perpendicular to the direction of  front propagation takes milliseconds, whereas once this has happened, the row itself fills in microseconds ({\it ``zipping''}). 
Numerical simulations confirm this view and are in quantitative agreement with the experiments. 
\end{abstract}
\maketitle
Micro-structured materials can show a superhydrophobic behavior with effective contact angles of $160^\circ$ and beyond (``Lotus effect''). For many applications this effect is wanted: when a liquid droplet is deposited on micro-textured
 hydrophobic surfaces, it can bead off completely and wash off contaminations 
very efficiently \cite{Degennes03,Barth97,Bico99,Quere,Marmur04}.
Superhydrophobic materials  are now used for a wide series of applications in  medicine, coatings,  self-cleaning, textiles, and microfluidics  \cite{Lauga05,Squire05,Otten04,Erbil03}. However, under certain conditions, the superhydrophobicity (``Cassie-Baxter state'' \cite{Cassie44}, hereafter CB) spontaneously breaks down \cite{Bonn00,He03}: fluid enters in between the  micro-structures and spreads, resulting into a smaller contact angle  (``Wenzel state'' \cite{Wenzel36}, hereafter W). In many cases this transition from the CB to the W state is highly desirable . For example, water-repellent layers on  leaves keep plants healthy, but also complicate spraying crops  effectively with pesticides \cite{Klein00}. Other examples are  heterogeneous porous catalysts, where superhydrophobicity is an unwanted effect as it reduces the contact area \cite{cat}. 

The objective of this Letter is to characterize and analyze the dynamics of the spontaneous breakdown of superhydrophobicity in terms of geometry, wetting, liquid-gas surface tension $\sigma_{lg}$, and the liquid viscosity $\eta$.

The fabrication of highly precise and controllable 
micro-structured surfaces becomes possible through a
micro-molding technique \cite{Jong06}. As material,
we choose  a solution of styrene-butadiene-styrene 
(a block-copolymer commercially available as 
``kraton'') dissolved in toluene. Kraton is used because thin layers are easily removable from the mold.
The thickness of the resulting translucent film is only $\sim$ 40 $\mu m$,
allowing for optical imaging of the CB to W transition
through the bottom of the film.

\begin{figure}[t!]
\begin{center}
(a)
\includegraphics[width=3cm,keepaspectratio]{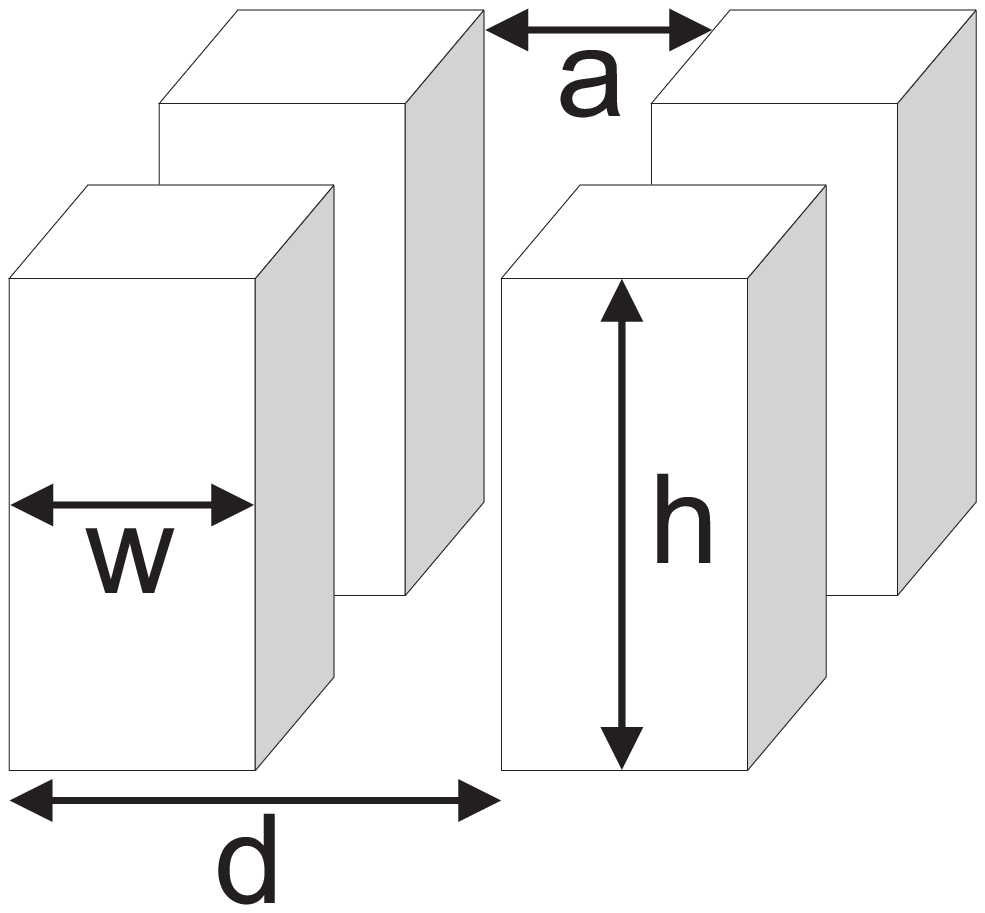}
\hspace{0.5cm}
(b)
\includegraphics[width=3.6cm,keepaspectratio]{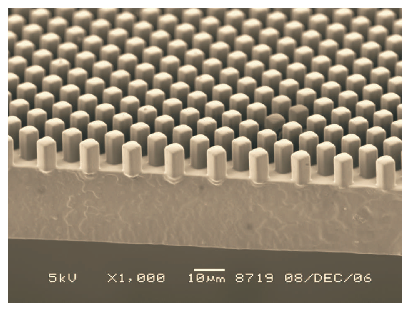}
\caption{Sketch (a) and scanning electron microscopy (SEM) picture (b) of the micro-patterned substrate.  The geometrical dimensions of the pattern are 
the pillar height $h=10 \mu m$, the pillar width $w=5\mu m$ (both kept
fixed for all experiments), and the gap width  
$a$, which is varied between $2\mu m$ and $17\mu m$.
The wavelength $d$ of the pattern is $d= a + w$.
In (b) we have $a= 5\mu m$. Typical equilibrium contact angles 
for a smooth surface made up of the same material are $\theta\sim  
100^{\circ}$.} 
\label{fig:2}
\end{center}
\end{figure}

For the experiments presented here we choose a 
regular,  periodic  structure of square pillars with height $h=10\mu m$,
width $w= 5\mu m$ (both kept fixed for all our experiments),  
and as control parameter 
a gap width $a$ between  $2\mu m$ and $17 \mu m$.
The wavelength $d=w+a$ of the regular square lattice is  thus  between $7\mu m$ and $22 \mu m$ (see figure \ref{fig:2}). 

Water droplets are deposited on the surface with a syringe pump at a very low flow rate ($5 \mu$l/min) from an outlet of a vertical thin tube
 (outer diameter 0.158 $mm$). The outlet is set parallel to the flat film, 2 $mm$ above it.  A typical droplet grows slowly and reaches the dry surface within a minute. Then the flow is stopped and a (meta-)stable 
drop in the CB state (contact area $\sim$ 1 $mm^2$) is observed. 
For the chosen micro-textured  surfaces the (effective)  contact angle in the CB state  is $\theta\sim$ $160^{\circ}$, while for a smooth surface made  of the same material it is $\theta\sim  100^{\circ}$. 

\begin{figure}[t!]
\begin{center}
\includegraphics[width=8.0cm,keepaspectratio]{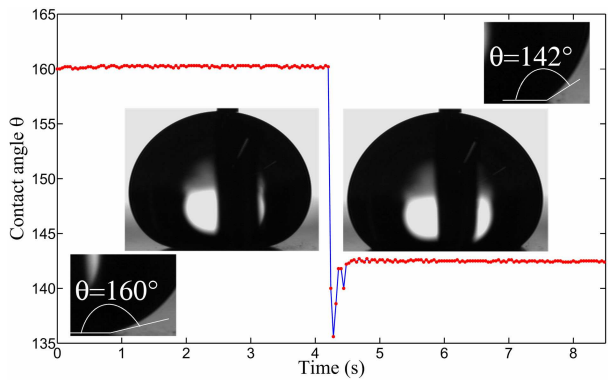}
\caption{Transition from the (meta-)stable CB to the W state. A drop softly deposited on a micro-patterned surface can stay suspended with air pockets trapped 
in the grooves underneath the liquid (left). At some point the CB state spontaneously breaks down. The drop then homogeneously wets the substrate, resulting in a lower contact angle (right).} 
\label{fig:1}
\end{center}
\end{figure}

At some point the CB state can spontaneously break down (figure \ref{fig:1} ). 

\begin{figure}[t!]
\begin{center}
\includegraphics[width=8.5cm,keepaspectratio]{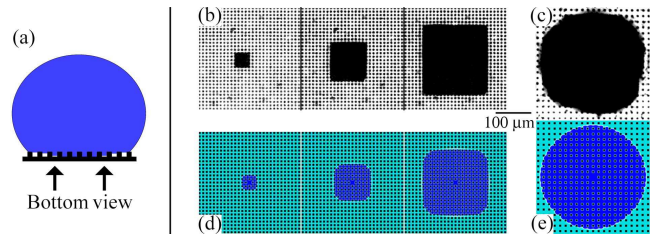}
\caption{(color) Bottom views of the front evolution of the  transition,
as sketched in (a). In (b)   three snapshots for the case 
with $a=5 \mu m$ are shown, leading to square-shaped wetted area. In (c) it is $a=11 \mu m$, resulting in a circular wetted area. Figures (d) and (e) show the results of the corresponding numerical simulations with the Lattice Boltzmann method  with $a= 5 \mu m$ and $11\mu m$, respectively. The infiltration point has been centered in the figure but is not located in the center of the droplet's base: its dependence on local energy barriers can indeed produce uncertainty in its location. } 
\label{fig:3}
\end{center}
\end{figure}

The extremely rapid filling process towards the W state
is recorded from the bottom of the film 
(see figure \ref{fig:3}a for a sketch)
with a microscope and a high-speed imaging system (Photron Ultima APX-RS), using  frame rates  in the range between 4000-50000 fps.
In figure \ref{fig:3} the {\it large scale} time evolution 
of the filling process is shown. After fluid sinks down at a certain point, a lateral spreading develops which sensitively depends on the parameters characterizing the micro-textured surface: while for large gap width $a=11 \mu m$ a round shape  of the fully wetted area is emerging  (figure \ref{fig:3}c), 
at smaller $a=5 \mu m$  the propagating fronts
reflect  the structure of the underlying square lattice,
leading to a square-shaped wetted area 
(figure \ref{fig:3}b). At even smaller $a=2 \mu m$ 
we never observed any transition to the W state.

\begin{figure}[t!]
\begin{center}
(a)\includegraphics[width=6.0cm,keepaspectratio]{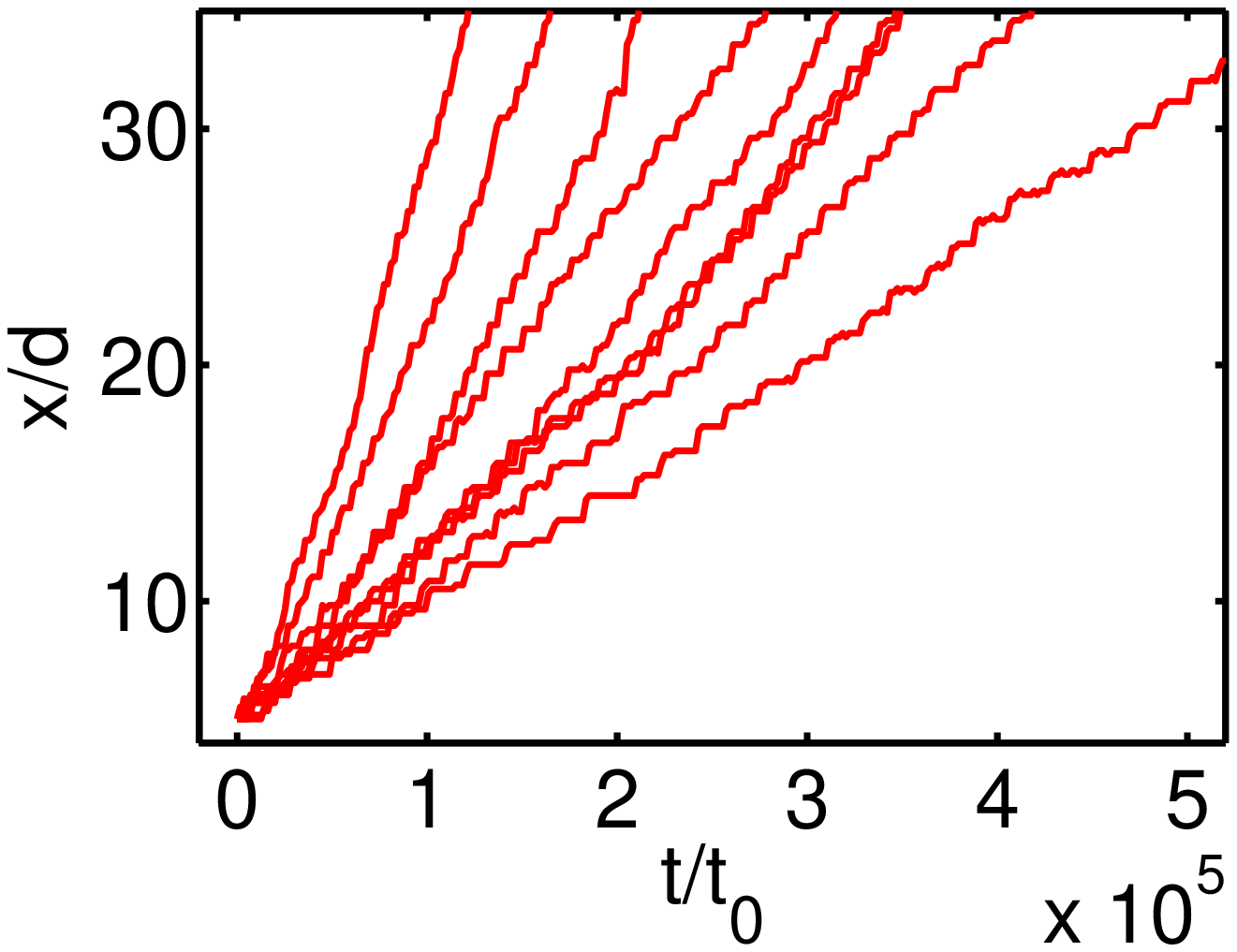}
(b)\includegraphics[width=6.0cm,keepaspectratio]{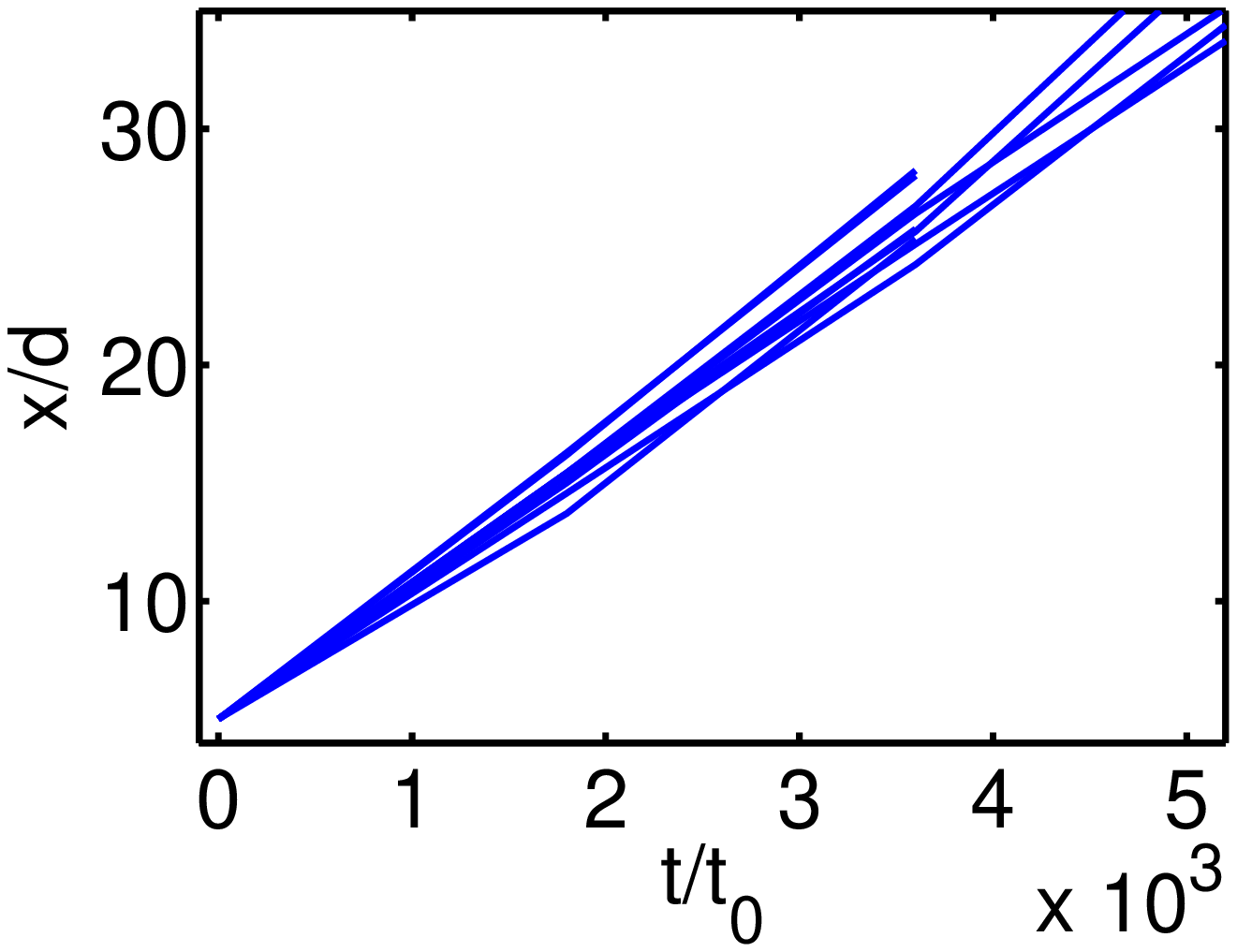}
\caption{Wetting front evolution for micro-patterned surfaces with
different gap-width $a$. We show a case where $a=5 \mu m$ (a) and a case where $a=11 \mu m$ (b). In both cases, nine different experiments were
 carried out with the same sample. The position of the 
front has been normalized with respect to the pattern wavelength 
$d=a+w$ and time has been made dimensionless with $t_{0}$ as in equation 
(\ref{eq:time}). Note the difference in front 
speeds $v$ in the two cases: in the small wavelength 
case (left) a {\it critical slowing down} ($v \sim$ 7 $mm/s$) 
induces large dispersion and {\it zipping} is revealed. The 
large wavelength case (right) is less dispersed and much faster, 
reaching velocities of $v \sim 7 \cdot 10^{2} mm/s$. } 
\label{fig:4}
\end{center}
\end{figure}

For the wetting cases, the velocity of front propagation
drastically depends on the gap width $a$ (see figure \ref{fig:4}): 
For  $a=11 \mu m$ we measure a mean velocity $v\sim  700 $mm/s, two  orders of magnitude faster than the case with $a=5 \mu m$, where $v \sim $ 7 $mm/s$.  Moreover, in the latter case the front propagates in a stepwise manner (figure \ref{fig:4}a) and fronts at different positions on the sample or in different directions show a remarkable variation in their velocity (``dispersion''). 

\begin{figure}[t!]
\begin{center}
(a)
\includegraphics[width=5cm,keepaspectratio]{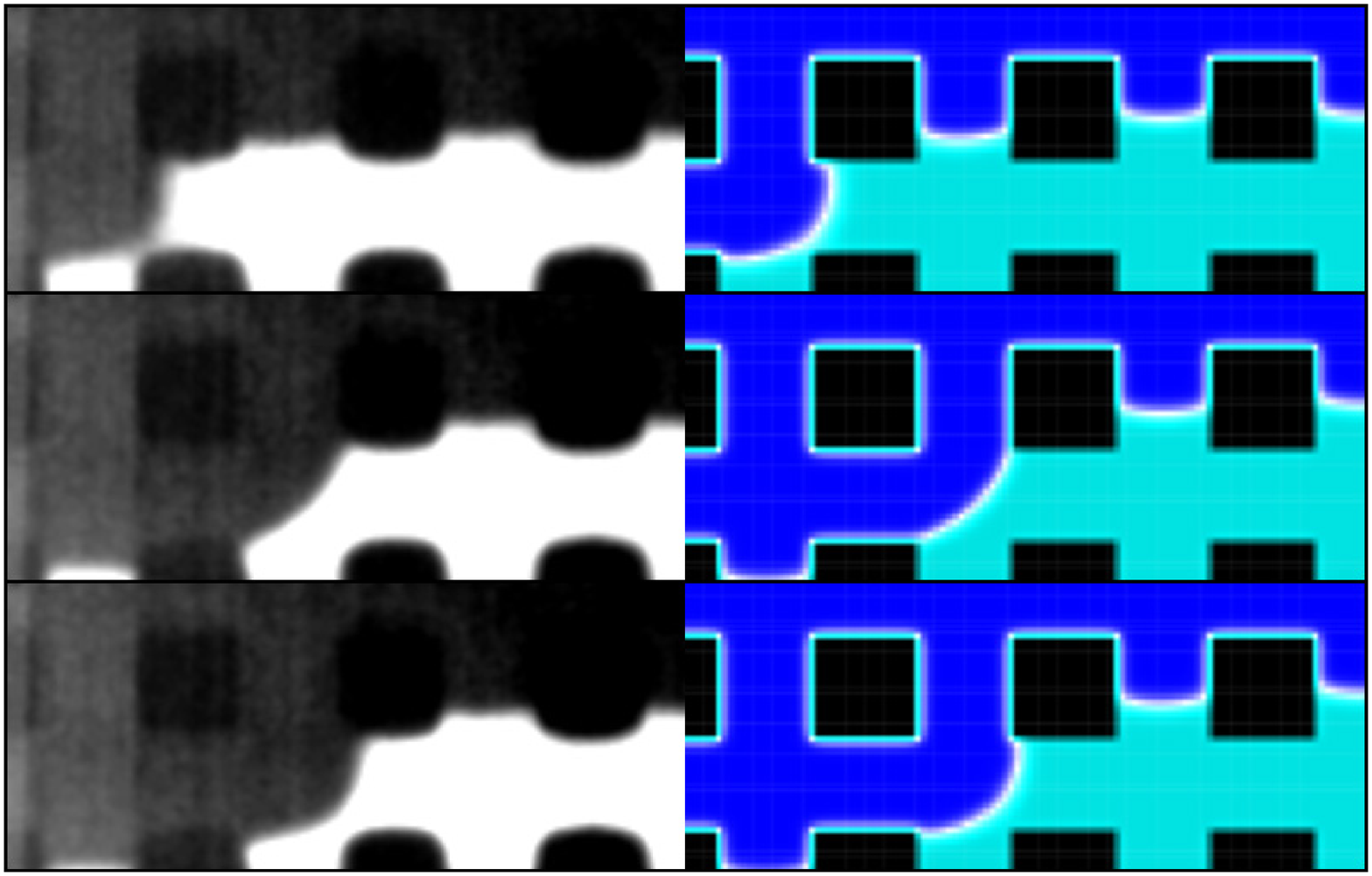}
(b) \hspace{0.2cm} \\
(c)
\includegraphics[width=6cm,keepaspectratio]{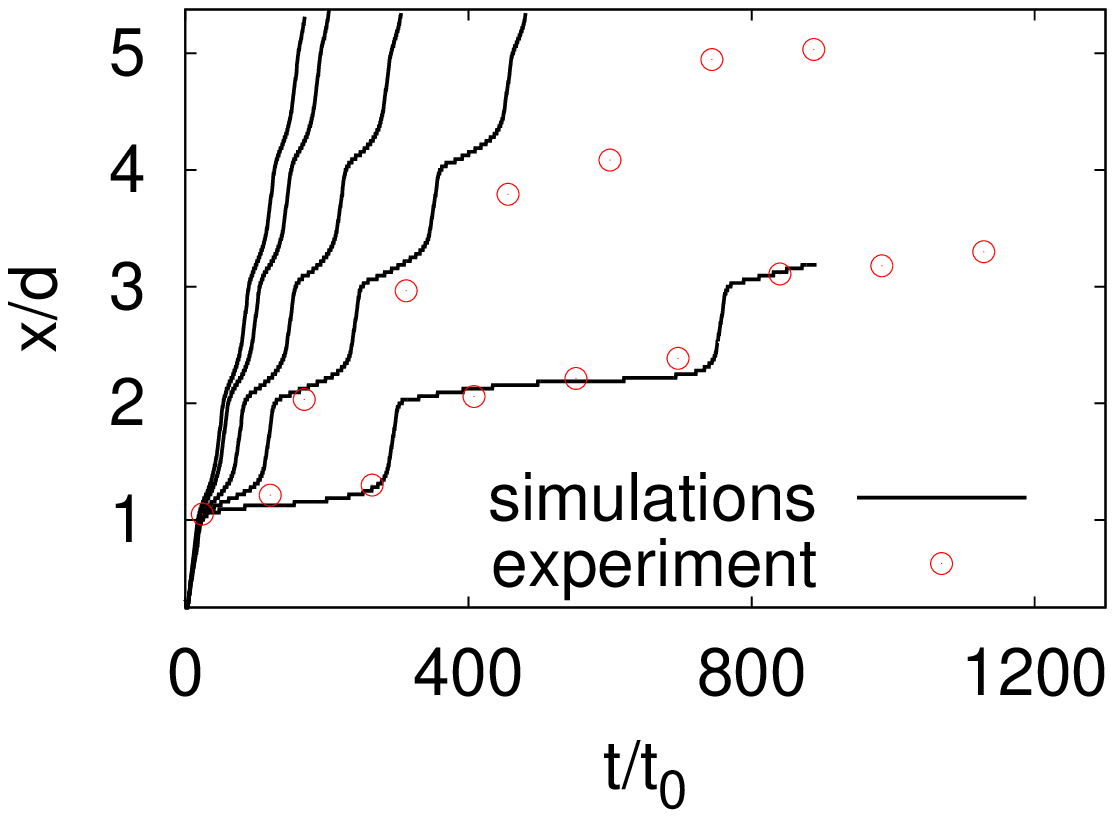}
\caption{The lateral zipping mechanism (left to right) close to the critical point ($\theta \sim \theta_{c}$). (a) Snapshots of the zipping mechanism 
recorded in an experiment with $a=5 \mu m$ with the front propagation from top to bottom. (b) The results are also compared with snapshots from numerical simulations. (c) Numerical results for the front propagation
(lines) in the lateral zipping process close to the critical point.
 From top to bottom we show the following cases: 
$\cos \theta/\cos \theta_{c} \sim 0.868, 0.890, 0.927, 0.945, 0.971$. 
The geometrical aspect ratio for the simulations is chosen to be the 
same as in the experiment with $a=5 \mu m$, whose typical outcomes are also reported ($\circ$).} 
\label{fig:5}
\end{center}
\end{figure}

For the more striking case of the square-shaped wetted area
($a=5 \mu m$) we show details of the wetting dynamics in 
figure \ref{fig:5}a. The front is here advancing from top to bottom and is slowing down  while wetting occurs in a {\it zipping} manner:  the timescale of the front for entering a new row is slow as already shown above (figure \ref{fig:4}a), namely typically $d/v \sim 1.4ms$,  the timescale for  the filling of the once entered row itself is about two orders of  magnitude faster, $\sim 0.01 ms$.  

In order to better reveal the physics of the wetting
mechanism, we performed numerical simulations with 
a three-dimensional Lattice Boltzmann algorithm  \cite{BSV92} for single component multiphase fluids. Wetting properties with surface tension are 
introduced as explained in \cite{LBM06a},
leading to wetting angles comparable with those in the experiments.
Geometrical structures are reproduced with the same aspect ratio as 
in the experiment.  Similarly, in the large gap width
 case $a=11 \mu m$ (nearly) spherical wetted areas are observed 
(figure \ref{fig:3}e), whereas in the case $a=5 \mu m$ we observe square-shaped wetted areas, see figure \ref{fig:3}d.
Also the zipping wetting behavior is reproduced in these simulations
for $a=5 \mu m$ (see figure \ref{fig:5}b).
When repeating the  simulations with a gap width  $a=2 \mu m$,  we 
do not observe any lateral infiltration of the substrate
and the CB state remains stable, again, just as experimentally observed. 
How to explain the transition from CB to W ? The energy of a droplet in equilibrium on a substrate is monotonically increasing with the effective contact angle \cite{He03}. As a result, when the CB effective contact angle is higher than the W one, we would argue that the CB droplet always collapses towards the W state. Anyhow, intermediate states with higher energy can be encountered \cite{He03}: they represent energy barriers to overcome and work has to be done in order to induce the transition \cite{He03,Bico99}. The bottom-view observations reveal that the micro-structured surface is not wetted all at once, but the infiltration starts locally.  In our case, this happens spontaneously after a few seconds whereas in other cases (i.e. larger $h$)  the infiltration can take longer and one may want to trigger it.  Once the transition has started, how to theoretically understand  the shape of the wetted area, the zipping, and the different involved timescales ?
For the dynamics of  liquid moving in between the posts, both the free surface and wall interactions play a role. A small advancing displacement $dx \ll a$ of the interface within the posts would take place in favour of a gain in surface energy due to the reduction of liquid-gas interface ($\sigma_{lg} a  dx  $) on the top of the pillars. This is a  {\it pulling} mechanism for the transition that must be balanced with the energy {\it loss} to wet a small portion of hydrophobic wall ($\sigma_{lg}  (2h+a) dx  \cos \theta$,  $\cos \theta <0$).  The overall energy gain is 
\be\label{eq:gainbare}
d E_{s}=\sigma_{lg} a dx +\sigma_{lg} (2h+a) dx   \cos \theta.
\ee
The limit $d E_{s} \rightarrow 0$ characterizes a critical contact angle $\theta_{c}$ for the wetting properties above which the liquid can spread horizontally through the posts. For square posts and square geometry it is equal to
\be \label{thetac}
\cos \theta_{c}=-1+\frac{2h}{2h+a}.
\ee
Note that in this equation there is no reference to the width $w$ of the posts because we describe the horizontal filling, different from a simultaneous vertical collapse \cite{Degennes03,He03}.
With this critical value, the energy gain (\ref{eq:gainbare}) can be rewritten as
\be\label{eq:gain}
d E_{s}=\sigma_{lg} a dx \left( 1-\frac{\cos \theta}{\cos \theta_{c}} \right)
\ee
where for $\cos \theta < \cos \theta_{c}$ no favorable propagation is expected  while in the other case, when $\cos \theta_{c} < \cos \theta <0$, the propagation is energetically favored. For the two analyzed cases $a=5 \mu m$ and  $11 \mu m$ we obtain $\theta_c = 101.54^{\circ}$ and $110.78^{\circ}$, respectively.  As $\theta = 100^{\circ}$ for water on flat kraton,  we thus understand that for gap widths smaller than $a = 4.2 \mu m$ the propagation to the W state is energetically not possible.  

We can also understand the {\it critical slowing down} of the front close to the critical angle  $\theta_{c}$ (figure \ref{fig:4}). Therefore we have to identify the relevant time scale $\tau$ for the  system and quantify its fluctuations with respect to the  geometry. To do so, we estimate the energy cost in  terms of dissipation and compare it with the change in surface energy of equation (\ref{eq:gain}). With the small dimensions under 
consideration, flows can be regarded as laminar and viscous 
dissipation is dominating as compared  to flow inertia. A simple
 estimate for the rate of viscous dissipation per unit volume
 in a  fluid with viscosity $\eta$ is $\epsilon \approx \eta 
\dot{\gamma}^2$ \cite{Landau59}, with $\dot{\gamma}\sim 1/\tau$
 the characteristic shear rate. If we integrate  it over  the small
 volume $dV=a h dx$ and  time lag $\tau$, we obtain the viscous 
contribution for the energy as $ d E_{d}\approx \epsilon \tau dV = \eta a h dx/{\tau}$.
Balancing $d E_{s}$ in equation (\ref{eq:gain}) with $d E_{d}$,
we deduce the time scale 
\be
\label{eq:time}
\tau={t_{0}\over (1-\frac{\cos \theta}{\cos \theta_{c}}) }
\ee
with $t_{0}= \eta h/\sigma_{lg} 
\sim 10^{-7}s$ for typical values of $\eta$, $\sigma_{lg}$, and $h=10 \mu m$. Equation (\ref{eq:time}) reveals the origin of the critical  slowing down of the front propagation, its dispersion and the appearance of zipping: if $\theta \sim \theta_{c}$, the timescale $\tau$  of front propagation {\it diverges} as $\tau \sim {1 / |\theta-\theta_{c}|}$. Around this 
{\it critical point}, a small uncontrollable variation 
(imperfections of the microstructure, dust deposit) 
in the local wetting angle translates into a 
huge dispersion in the time scales, which is consistent with 
the experimental observation shown in figure \ref{fig:4}a. A flat
 front is slowly allowed to proceed further, and entering a new 
row perpendicular to the direction of the front can  take up 
to milliseconds, whereas once this has happened, the row itself 
is filled on  a faster time scale ({\it zipping}).
 In this way a square-shaped propagating wetting pattern emerges 
(see the case with $a=5 \mu m$ in figure \ref{fig:3}b
 where $\theta_{c}\sim 101.54^{\circ}$ and $\theta\sim 100^{\circ}$). 

On the contrary, when approaching the other limit $\cos\theta/\cos\theta_{c} \ll 1$ (as  in figure \ref{fig:3}c, where $\theta\sim 100^{\circ}$ and  $a=11 \mu m$, implying  $\theta_{c}\sim 110.78^{\circ}$) the dynamics becomes more and more determined by the time scale $t_{0}$ itself (see figure \ref{fig:4}b) and the propagation through the posts is expected to be smooth. In this limit it is the pure pulling mechanism of surface tension in equation (\ref{eq:gain}) that dominates  the  spreading dynamics and is prevalent over wall effects \cite{Ferimigier03}. For this reason, the geometrical properties of the lattice do not emerge and the front assumes an almost circular shape.

In conclusion, we have experimentally, numerically, and theoretically revealed the origin of zipping wetting behavior at the spontaneous breakdown of superhydrophobicity.  We observed that the wetting process starts locally from a single point and then proceeds laterally, depending on a critical contact angle.

Close to this critical point the driving energy for the lateral filling reduces, the front propagation slows down through viscous effects and zipping wetting is observed. As a consequence, the front propagates in a stepwise manner and square-shaped \cite{Wulff}  wetted areas emerge.  Although we have observed the origin of this zipping process, its dynamical details have not been characterized here and will be the object of a forthcoming paper \cite{Pirat07}.

The critical contact angle (equation (\ref{thetac})) can be calculated from the geometrical properties of the micro-textured material and is consistent with our experimental and numerical findings. 

Our results are useful for the design and fabrication of micro-structured surfaces with certain wetting properties: for given hydrophobic  material and liquid, the post height and the gap width can simply be calculated to obtain some desired critical contact angle. 

The next step will be to control the time scale of the local initiation point, characterize the dynamics of the zipping process and extend our analysis to more  complicated micro-patterned structures, including three-dimensional ones .

We gratefully acknowledge discussions with
 B.\ Andreotti, 
L.\ Courbin,
L.\ Lefferts,
F.\ Mugele, and 
A.\ Prosperetti.   
M.S.\ and B.M.B.\ 
thank
STW (Nanoned Programme) and A.M.P.\ thanks
Microned for financial support.

\end{document}